\documentclass[pra,twocolumn,superscriptaddress,showpacs]{revtex4}
\usepackage{CJK}   
\usepackage{indentfirst}  
\usepackage{bm}    
\usepackage{graphicx}  
\usepackage{epsfig} 
\usepackage{amsmath} 
\usepackage{amssymb}
\usepackage{bm,hyperref}

\begin{document}
\begin{CJK*}{UTF8}{gbsn}
\title{Superfluidity of Bose-Einstein condensates in ultracold atomic gases}
\author{Qizhong Zhu(朱起忠)}
\affiliation{International Center for Quantum Materials, School of Physics, Peking University, Beijing 100871, China}
\author{Biao Wu(吴飙)}
\affiliation{International Center for Quantum Materials,  School of Physics, Peking University, Beijing 100871, China}
\affiliation{Collaborative Innovation Center of Quantum Matter, Peking University, Beijing 100871, China}
\date{\today}

\begin{abstract}
Liquid helium 4 had been  the only bosonic superfluid available in experiments
for a long time. This situation was changed in 1995, when  a new superfluid
was born with the realization of the Bose-Einstein 
condensation in ultracold atomic gases.  The liquid helium 4 is
strongly interacting and has no spin;  there is almost no way to change its parameters, 
such as interaction strength and density. The new superfluid, Bose-Einstein condensate (BEC), 
offers various aspects of advantages over liquid helium. 
On the one hand, BEC is weakly interacting and has spin degrees of freedom. 
On the other hand, it is convenient to tune almost all the parameters of a BEC, 
for example, the kinetic energy by spin-orbit coupling, the density by 
the external potential, and the interaction by Feshbach resonance.
Great efforts have been devoted to studying these new aspects of superfluidity,
and the results have greatly enriched our understanding of superfluidity.
Here we review these developments by focusing on the stability and critical velocity of
various superfluids. The BEC systems considered include a uniform superfluid in free space, 
a superfluid with its density periodically modulated, a superfluid with 
artificially engineered spin-orbit coupling, and a superfluid of pure spin current.
Due to the weak interaction, these BEC systems can be well described by  
the mean field Gross-Pitaevskii theory and their superfluidity, in particular 
critical velocities, can be examined with Landau's theory of superfluid. 
Experimental proposals to observe these new aspects of superfluidity
are discussed.
\end{abstract}
\maketitle

\section{Introduction} 
Superfluidity, as a remarkable macroscopic quantum phenomenon,
was first discovered in the study of liquid helium 4 in the 1938~\cite{kapitsa,allen}.
Although it is found theoretically that superfluidity is a general phenomenon 
for interacting boson systems~\cite{Nozieres}, liquid helium had been 
the only bosonic superfluid available in experiments until 1995. In this year, 
thanks to the advance of laser cooling of atoms, the Bose-Einstein condensation of 
dilute alkali atomic gases was realized experimentally~\cite{bec}; a new 
superfluid, Bose-Einstein condensate (BEC), was born. This addition to the family 
of superfluids is highly non-trivial as BECs offer various aspects of advantages 
over liquid helium that can greatly enrich our understanding of superfluidity. 

Great deal of work has been done to explore the properties of liquid helium 
as superfluid~\cite{helium}. 
However, these efforts have been hindered by limitations of liquid helium. As a liquid,
the superfluid helium is strongly-interacting system, which makes the theoretical description
difficult. At the same time, no system parameters, such as density and interaction strength,
can be tuned experimentally. And helium 4 has no spin degrees of freedom. 

BECs are strikingly different.  Almost all the parameters of a BEC can be controlled easily in
experiments: its kinetic energy, density, and the interaction between atoms can all be tuned
easily by engineering the atom-laser interaction, magnetic or optical traps, 
and the Feshbach resonance ~\cite{feshbach}. In addition, by choosing the atomic species and
using optical trap to release the spin degrees of freedom, one can also realize 
various types of new superfluids, including the spinor superfluid ~\cite{spinor} 
and the dipolar superfluid ~\cite{cr}. All these  are impossible with liquid helium. 
Moreover, as most ultracold gases are dilute and weakly-interacting,
controllable theoretical methods are available  to study these superfluids in detail.

In this review we mainly discuss three types of bosonic superfluids: 
superfluid with periodic density, superfluid with spin-orbit coupling, and superfluid of  pure spin current.
The focus is on the stability and critical velocities of various superfluids.
For a uniform superfluid, when its speed exceeds a critical value, the system suffers
Landau instability and superfluidity is lost. When the superfluid moves in a periodic potential,
with large enough quasi-momentum, new mechanism of instability, i.e., dynamical instability comes in. This instability usually dominates the Landau instability as it occurs 
on a much faster time scale. The periodic density also brings another twist: the superfluidity
can be tested in two different ways, which yield two different critical velocities. 
For a superfluid with spin-orbit coupling, a dramatic change is brought 
in, namely the breakdown of Galilean invariance. As a result, its
critical velocity will depend on the reference frame. The stability of a pure spin current 
is also quite striking. We find that the pure spin current in general is not a super-flow.
However, it can be stabilized to become a super-flow with 
quadratic Zeeman effect or spin-orbit coupling. Related experimental proposals
are discussed. 

The paper is organized as follows. In Sec. \ref{basics}, we discuss 
briefly the basic concepts related to the understanding of superfluidity, including 
Landau's theory of superfluidity, mean field
Gross-Pitaevskii equation, and Bogoliubov excitations. These concepts are illustrated
with the special case of a uniform superfluid. We then apply these general
methods to study in detail the superfluid in periodic potentials in Sec. \ref{lattice}, the superfluid
with artificially engineered spin-orbit coupling in Sec. \ref{socsuperfluid}, and finally the superfluid of pure spin current in Sec. \ref{spincurrent}. In these three superfluids, special attention is paid
to their excitations, stabilities, and  critical velocities. 
We finally summarize in Sec. \ref{summary}.

\section{Basic concepts of superfluidity}\label{basics}  
\subsection{Landau's theory of superfluidity}
Superfluid is a special kind of fluid which does not suffer dissipation 
when flowing through a tube. It loses its superfluidity only 
when its speed exceeds a certain critical value. 
The superfluidity of liquid helium 4 was first explained by L. D. Landau ~\cite{landau}. 
He considered a superfluid moving inside a stationary tube with velocity $\mathbf{v}$. 
Since the system is invariant under the Galilean transformation, this scenario is 
equivalent to a stationary fluid inside a moving tube.
If the elementary excitation in a stationary superfluid with momentum $\mathbf{q}$ 
has energy $\epsilon_0(\mathbf{q})$,  then the energy of the same excitation in 
the background of a moving fluid with $\mathbf{v}$ is 
$\epsilon_{\mathbf{v}}(\mathbf{q})=\epsilon_0(\mathbf{q})+\mathbf{v}\cdot\mathbf{q}$.
A fluid experiences friction only through emitting elementary excitations, and it is a superfluid
if these elementary excitations are energetically unfavorable. In other words, a superfluid satisfies
the constraint $\epsilon_{\mathbf{v}}(\mathbf{q})>0$. It readily leads to the well-known
Landau's criterion for superfluid, 
\begin{equation}
v<v_{\mathrm{c}}=\left(\frac{\epsilon_0(\mathbf{q})}{|\mathbf{q}|}\right)_{\mathrm{min}}.
\label{min}
\end{equation}
Here $v_{\mathrm{c}}$ is the critical velocity of the superfluid, which is determined by the smallest
slope of the excitation spectrum of a stationary superfluid. 

Another way of deriving the formula of critical velocity is from the point view of Cerenkov radiation.
Consider a macroscopic impurity moving in the superfluid generates an excitation.
According to the conservations of both momentum and energy, we should have
\begin{eqnarray}  \label{co1}
m_0{\mathbf{v}}_i&=&m_0{\mathbf{v}}_f+{\mathbf{q}}\,, \\
\frac{m_0{\mathbf{v}}_i^2}{2}&=&\frac{m_0{\mathbf{v}}_f^2}{2}+\epsilon_0({%
\mathbf{q}})\,,  \label{co2}
\end{eqnarray}
where $m_0$ is the mass of the impurity, ${\mathbf{v}}_i$ and ${\mathbf{v}}_f
$ are the initial and final velocities of the impurity, respectively. 
The above two conservations (\ref{co1}) and (\ref{co2}) can not be satisfied simultaneously
when
\begin{equation}
v\approx|{\mathbf{v}}_i|\approx|{\mathbf{v}}_f|<v_{\rm c}=\left(\frac{\epsilon_0({\mathbf{q}})}{|{\mathbf{q}}|}\right)_{\rm min}\,.
\label{cerenkov}
\end{equation}
The critical velocity $v_{\rm c}$ here has the same expression as that obtained from Landau's criterion.
If the excitations are phonons, i.e., $\epsilon_0({\mathbf{q}})=c|{%
\mathbf{q}}|$, then $v_{\rm c}<c$. This means that
the impurity could not generate phonons in the superfluid and would not
experience any viscosity when its speed was smaller than the sound speed.
This is in fact nothing but the Cerenkov radiation~\cite{key2,key3}, where a
charged particle radiates only when its speed exceeds the speed of light in
the medium.

\begin{figure}[!h]
\includegraphics[width=8cm]{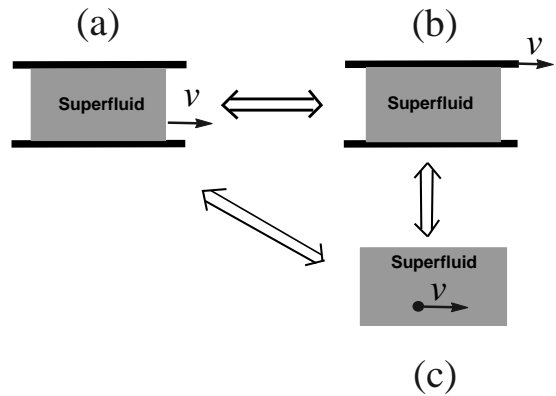}\newline
\caption{(a) A superfluid moves inside a stationary tube.
(b) The superfluid is dragged by a tube moving at the speed of $v$.
(c) An impurity moves at $v$ in the superfluid. The two-way arrow indicates the equivalence between different
scenarios.}
\label{GI}
\end{figure}

These two different ways of derivation are equivalent when the system has the Galilean invariance.
By transforming to another reference frame illustrated in Fig. \ref{GI}(b), 
the superfluid can be viewed as being dragged by a moving tube. 
We then replace the moving tube with a macroscopic impurity moving
inside the superfluid as shown in Fig. \ref{GI}(c). Consequently,
for systems with Galilean invariance, the critical velocity of a superfluid moving inside a tube
without experiencing friction is just the critical velocity of an impurity moving in
a superfluid without generating excitations. For systems without the Galilean invariance,
such an equivalence is lost as we will see later with the spin-orbit coupled BEC.

For ultracold bosonic gases,
the low energy excitation is phonon, linear with respect to momentum $\mathbf{q}$,
thus the critical speed of this superfluid is just the sound speed. 
Taking into account the non-uniformity of the trapped gas, the critical velocity measured in
experiments agrees well with the value predicted by Landau's theory~\cite{cv}.
For liquid helium 4, as there is another
kind of elementary excitations called rotons, the critical speed of superfluid helium 4 is largely determined
by the roton excitation, much smaller than its sound speed. Nevertheless, the critical velocity
measured in experiments is still one order of magnitude smaller than the value predicted by the 
theory~\cite{He_exp}.
So it is remarkable that Landau's prediction of critical velocity was experimentally 
confirmed  with BEC almost six decades after its invention.

One remark is warranted on  Landau's theory of superfluidity. 
Landau's criterion (\ref{min}) of critical speed does not apply for many 
superfluids. However, Landau's  energetic argument for  superfluidity is 
very general and can be applied to all the cases considered in this review. 
We shall use this argument to determine the critical speeds of various superfluids. 

\subsection{Gross-Pitaevskii equation and Bogoliubov excitation} \label{gp}
As liquid helium 4 is a strongly interacting system, the calculation of its excitation spectrum theoretically
is challenging. However, for a dilute ultracold bosonic gases, convenient yet precise approximations can
be made to
determine its excitation in theory. Due to the dilute nature and short range interaction, the interaction
between atoms can be approximated by a contact interaction. Furthermore, for low energy scattering of
bosonic atoms, the $s$-wave scattering channel dominates. So, as a good approximation, 
the actual complex interaction between
atoms is replaced by an effective $s$-wave contact interaction, i.e., 
$U(\mathbf{r}_1,\mathbf{r}_2)=4\pi\hbar^2 a\delta(\mathbf{r}_1-\mathbf{r}_2)/M$, where $a$ is the $s$-wave
scattering length. At zero temperature, assume all the particles condense into the same orbit $\psi(\mathbf{r})$,
then its evolution is governed by the mean field Gross-Pitaevskii (GP) equation~\cite{pethick},
\begin{equation}
i\hbar\frac{\partial\psi(\mathbf{r},t)}{\partial t}=-\frac{\hbar^2\nabla^2}{2M}\psi+V(\mathbf{r})\psi+c|\psi|^2\psi,
\end{equation}
where $V(\mathbf{r})$ is the trapping potential or other external potential, and $c=4\pi\hbar^2 a/M$ is the interaction parameter
or coupling constant. 
The GP energy functional reads
\begin{equation}
\mathcal{E}[\psi,\psi^*]=\int d\mathbf{r}\left[-\frac{\hbar^2}{2M}|\nabla\psi|^2+V(\mathbf{r})|\psi|^2
+\frac{c}{2}|\psi|^4\right].
\end{equation}
Note that here the wave function $\psi$ is normalized to the total particle number $N$.
The approximation we make here is the mean field approximation, and the interaction between
atoms is replaced by an effective mean field potential. The validity of the approximation usually
depends on the condensed fraction.
 For weakly interacting dilute bosonic gases near zero temperature,
the condensed fraction can be more than 90 percent. In these situations, the GP equation works well.
It has been used to calculate the collective excitations of trapped BEC, as well as vortex dynamics.
Pretty good agreement is achieved between theory and experiment.
With a stochastic term describing the effect of thermal atoms, the modified stochastic GP equation can also
simulate BEC systems at finite temperature~\cite{temp}.
In addition, for atoms with spin, one can also derive a multicomponent GP equation to describe a spinor
superfluid~\cite{uedarmp}. 

The stationary state the GP equation describes is a zeroth order approximation in some sense. It only
takes into account the interaction between the condensed particles.
The first order approximation is to take into account the interaction between the condensed
particles and un-condensed particles. This is settled by the Bogoliubov theory of elementary excitations~\cite{statphys}.
The standard Bogoliubov theory is to diagonalize the mean field Hamiltonian by the Bogoliubov transformation.
Here we adopt an equivalent yet more convenient way to deal with this problem in inhomogeneous systems.
For a stationary state $\psi(\mathbf{r},t)$ that satisfies the GP equation, we consider some small time dependent 
perturbations $\delta\psi(\mathbf{r},t)$ 
added to this stationary state. The perturbed state $\Psi=\psi+\delta\psi$ also satisfies the GP equation,
\begin{equation}
i\hbar\frac{\partial\Psi(\mathbf{r},t)}{\partial t}=-\frac{\hbar^2\nabla^2}{2M}\Psi+V(\mathbf{r})\Psi+c|\Psi|^2\Psi.
\end{equation}
Expanding this equation to first order in the perturbation $\delta\psi$, one arrives at the equation of motion
of the perturbation $\delta\psi(\mathbf{r},t)$,
\begin{eqnarray}
i\hbar\frac{\partial\delta\psi}{\partial t}&=&-\frac{\hbar^2}{2M}\nabla^2\delta\psi
+V(\mathbf{r})\delta\psi \notag \\
&&+2c|\psi|^2\delta\psi+c\psi^2\delta\psi^*,
\end{eqnarray}
and its complex conjugate partner.
Assume the state before perturbation is a stationary state with the wave function 
$\psi(\mathbf{r},t)=\sqrt{n(\mathbf{r})}e^{-i\mu t/\hbar}$, and write the perturbations as 
$\delta\psi(\mathbf{r},t)=\sqrt{n(\mathbf{r})}e^{-i\mu t/\hbar}[u(\mathbf{r})e^{-i\omega t}-v^*(\mathbf{r})
e^{i\omega t}]$,  then we arrive at a pair of equations that $u(\mathbf{r})$ and $v(\mathbf{r})$ satisfy,
\begin{equation}
\mathcal{M}\left(\begin{array}{c}u(\mathbf{r}) \\v(\mathbf{r})\end{array}\right)=\hbar\omega
\left(\begin{array}{c}u(\mathbf{r}) \\v(\mathbf{r})\end{array}\right),
\end{equation}
where the matrix $\mathcal{M}$ is given by
\begin{equation}
\mathcal{M}=\left(\begin{array}{cc}
H_0 & -cn(\mathbf{r}) \\
cn(\mathbf{r}) & -H_0
\end{array}\right),
\end{equation}
with $H_0=-\frac{\hbar^2}{2M}\nabla^2+V(\mathbf{r})-\mu+2cn(\mathbf{r})$.
By diagonalizing $\mathcal{M}$, one obtains two sets of solutions, but only the solution satisfying the constraint
$\int d{\bf r}\left(|u(\mathbf{r})|^2-|v(\mathbf{r})|^2\right)>0$ represents physical excitations. Note that since the characteristic matrix $\mathcal{M}$ is not hermitian,
its eigenvalues are not necessarily real. If its eigenvalue $\epsilon=\hbar\omega$ has nonzero imaginary
part, the perturbation $\delta\psi(\mathbf{r},t)=\sqrt{n(\mathbf{r})}e^{-i\mu t/\hbar}[u(\mathbf{r})e^{-i\omega t}-v^*(\mathbf{r})e^{i\omega t}]$ will grow exponentially with time, signaling that the state before perturbation suffers
{\it dynamical instability}. If its eigenvalue is real but negative, elementary excitations associated with the perturbations 
 will be energetically favorable and superfluidity is lost, which is called {\it Landau instability}.
For the excitations of repulsive Bose gases without external potentials, the low energy excitations must be non-negative and gapless from
the Hugenholtz-Pines theorem~\cite{HPtheorem}.

As a specific example, we apply the above formalism to the uniform Bose gas, namely, putting $V(\mathbf{r})=0$.
The wave function before perturbation is just $\psi=\sqrt{n}e^{-i\mu t/\hbar}$, and the perturbation has this
form $u(\mathbf{r})=u_{\mathbf{q}}e^{i\mathbf{q}\cdot\mathbf{r}}/\sqrt{V}$, 
$v(\mathbf{r})=v_{\mathbf{q}}e^{i\mathbf{q}\cdot\mathbf{r}}/\sqrt{V}$. Plugging these wave functions into the above
equations, one immediately obtains the excitation spectrum for a uniform Bose gas,
\begin{equation}
\epsilon_{q}=\sqrt{\epsilon_{q}^0\left(\epsilon_{q}^0+2cn\right)},
\end{equation}
where $\epsilon_{q}^0=\hbar^2 q^2/2M$ is the single particle spectrum.
At long wavelength or small momentum $q$, the excitation has the asymptotic form $\epsilon_q\sim
q\sqrt{\hbar^2 nc/M}$. This is nothing but phonon excitation, and $\sqrt{nc/M}$ is just the speed of sound.
From Landau's theory of critical velocity, we conclude that the critical velocity of a uniform superfluid
Bose gas is just the sound speed.

The method of mean field approximation and Bogoliubov transformation is very general, and
applies in other more complicated situations.
In the following discussion, we use this method to study  three types
 of superfluids, superfluid in
a periodic potential, superfluid with spin-orbit coupling and superfluid of pure spin current.

\section{Periodic superfluid}\label{lattice}
It is hard to change the density of helium 4 as it is a liquid. 
In contrast, we can easily modulate the density of a BEC which is a gas. 
When we put a BEC in an optical lattice, we obtain a superfluid whose density
is periodically modulated. One can even further periodically modulate the interatomic interaction of the BEC 
with optical Feshbach resonance~\cite{ZhangShaoLiang}. 
Supersolid helium 4 may be also regarded as a periodic superfluid as it can be viewed
as some superfluid defects (most likely vacancies) flowing in a helium solid 
lattice~\cite{supersolid}.  In this section, we use a BEC in an optical lattice 
as an example to examine the properties  of a periodic superfluid. 
Compared to the uniform superfluid in free space, a new type of instability,
i.e., the dynamical instability is found when the quasi-momentum $k$ of the superfluid is larger than a
critical value. Usually the dynamical instability dominates the accompanying 
Landau instability as it happens on a much faster time scale~\cite{Fallani2004PRL}. 
The presence of the periodic potential also brings along another critical velocity.

\subsection{Stability phase diagram}
Now we study the superfluidity of a BEC in a periodic potential~\cite{wu1,wu2}, 
which is provided by the optical lattice in cold atom experiments. For simplicity, we consider
a quasi-one-dimensional BEC, confined in a cigar-shaped trap.
We treat the system with the mean field theory and obtain the grand-canonical GP Hamiltonian
\begin{equation}\label{hol}
\mathcal{H}=\int_{-\infty}^{\infty}dx\left[\psi^*\left(-\frac{1}{2}\frac{\partial^2}{\partial x^2}+v\cos x\right)\psi
+\frac{c}{2}|\psi|^4-\mu|\psi|^2\right],
\end{equation}
where all the variables are scaled to be dimensionless with respect to a set of characteristic parameters
of the system, the atomic mass $M$, the wave number $k_L$ of the laser light generating the optical lattice,
and the average density $n_0$ of the BEC. The chemical potential $\mu$ and the strength of the periodic
potential $v$ are in units of $4\hbar^2k_L^2/M$, the wave function $\psi$ is in units of $\sqrt{n_0}$,
$x$ is in units of $k_L/2$, and $t$ is in units of $M/4\hbar^2k_L^2$. The interaction constant is given by
$c=\pi n_0 a_s/k_L^2$, where $a_s>0$ is the $s$-wave scattering length.

For non-interacting case ($c=0$), diagonalizing the Hamiltonian will give the 
standard Bloch waves and energy bands. When the mean field interaction is 
turned on ($c\neq 0$), in principle the Hamiltonian allows for other types of solutions 
which have no counterpart in the non-interacting case~\cite{GapSoliton1,GapSoliton2}. 
Here we focus on the solutions which still have the
form of Bloch waves, i.e., $\psi_k(x)=e^{ikx}\phi_k(x)$, where $\phi_k(x)$ has the same period with the optical lattice.
$\phi_k(x)$ can be found by extremizing the Hamiltonian above~\cite{wu2}. 
The solution found in this way should satisfy the stationary GP equation with periodic potential,
\begin{equation}
-\frac{1}{2}\frac{\partial^2}{\partial x^2}\psi+v\cos x\psi+c|\psi|^2\psi=\mu\psi.
\end{equation}

To determine the superfluidity of these Bloch states, we must consider elementary 
excitations around these Bloch states, and check whether the excitation energy is 
always positive. Positive excitation energy indicates that the Bloch state is a local 
energy minimum, and it is stable against small perturbations. Due to the periodicity 
of the Bloch wave, the perturbations can be decomposed into different decoupled 
modes labeled by $q$,
\begin{equation}
\delta\phi_k(x,q)=u_k(x,q)e^{iqx}+v_k^*(x,q)e^{-iqx},
\end{equation}
where $q$ ranges between $-1/2$ and $1/2$ and the perturbation functions $u_k$ and $v_k$ are of 
periodicity of $2\pi$.

Following the similar method in Sec. \ref{gp}, we linearize the GP equation above to obtain the Bogoliubov equation that
$u_k$ and $v_k$ satisfy
\begin{equation}
\mathcal{M}_k(q)\left(\begin{array}{c}u_k \\v_k\end{array}\right)=\epsilon_k(q)\left(\begin{array}{c}u_k \\v_k\end{array}\right),
\end{equation}
where
\begin{equation}
\mathcal{M}_k(q)=\left(\begin{array}{cc}
\mathcal{L}(k+q) & -c\phi_k^2 \\
c\phi_k^{*2} & -\mathcal{L}(-k+q)\end{array}\right),
\end{equation}
with
\begin{equation}
\mathcal{L}(k)=-\frac{1}{2}\left(\frac{\partial}{\partial x}+ik\right)^2+v\cos x-\mu+2c|\phi_k|^2.
\end{equation}
This eigenvalue equation has two sets of solutions, one corresponds to physical excitations, 
which is mostly phonon excitation, and the other can be called anti-phonon that is not physical.
If the physical excitation $\epsilon_k(q)$ is positive, the Bloch wave $\psi_k$ is a local minimum,
and the system will have superfluidity. Otherwise, the system suffers Landau instability or dynamical
instability, depending on whether $\epsilon_k(q)$ is real negative or complex.

In the case of  free space $v=0$, the Bloch state $\psi_k$ becomes a plane wave
$e^{ikx}$. Then the operator $\mathcal{M}_k(q)$ becomes
\begin{equation}
\mathcal{M}_k(q)=\left(\begin{array}{cc}
q^2/2+kq+c & -c \\
c & -q^2/2+kq-c
\end{array}\right),
\end{equation}
and we recover the excitations  in the uniform case 
\begin{equation}\label{eq:epsilon}
\epsilon_{\pm}(q)=kq\pm\sqrt{cq^2+q^4/4}.
\end{equation}
One immediately sees that the excitation energy is always real, 
which means that the BEC flows in free space are always dynamically stable.

\begin{figure}
\begin{center}
\includegraphics[width=8cm]{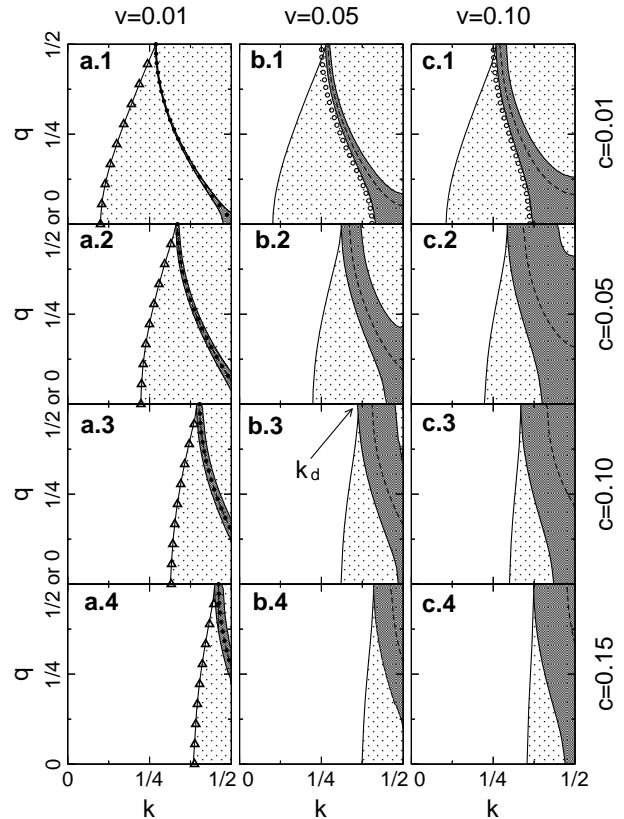}
\caption{Stability phase diagrams of BEC Bloch states in optical lattices. $k$ is the wave number 
of BEC Bloch waves; $q$ denotes the wave number of perturbation modes.
In the shaded (light or dark) area, the perturbation mode has negative 
excitation energy; in the dark shaded area, the mode grows or decays 
exponentially in time.  The triangles in (a.1-a.4) represent 
the boundary, $q^2/4+c=k^2$, of saddle point regions at $v=0$.  The solid
dots in the first column are from the analytical results of Eq.(\ref{eq:kqc}).
The circles in (b.1) and (c.1) are based on the analytical expression
(\ref{eq:bb}). The dashed lines indicate the most unstable modes for 
each Bloch wave $\psi_k$.}
\label{fig:kq}
\end{center}
\end{figure}

When there is periodic potential, the situation is dramatically different, where the 
excitation energy can have imaginary part, signaling the dynamical instability 
of the system. By numerically solving the Bogoliubov equation above, we show 
the stability phase diagrams for BEC Bloch waves in the panels of Fig. \ref{fig:kq}, 
where different values of $v$ and $c$ are considered.
The results have reflection symmetry in $k$ and $q$, 
so we only show the parameter region, $0 \le k \le 1/2$ and $0 \le q \le 1/2$. 
In the shaded area (light or dark) of each panel of Fig. \ref{fig:kq},
the excitation energy is negative, and the corresponding Bloch 
states $\psi_k$ are saddle points. For those values of $k$ outside the shaded 
area, the Bloch states are local energy minima and represent superfluids.   
The superfluid region expands with increasing atomic interaction $c$, and 
occupies the entire Brillouin zone for sufficiently large $c$.  
On the other hand, the lattice potential strength $v$ does not affect
the superfluid region very much as we see in each row.
The phase boundaries for $v\ll 1$ are well reproduced from the analytical 
expression $k=\sqrt{q^2/4+c}$ for $v=0$, which is plotted as triangles
in the first column. 

If $\epsilon_k(q)$ is complex, the system suffers dynamical instability, which is shown by the
dark-shaded areas in Fig. \ref{fig:kq}. 
The dynamical instability is the result of the resonance coupling
between a phonon mode and an anti-phonon mode by first-order
Bragg scattering. The matrix $\mathcal{M}_k(q)$
is real in the momentum representation, meaning that its complex eigenvalue
can appear only in conjugate pairs and they must come from a pair of real
eigenvalues that are degenerate prior to the coupling. Degeneracies or resonances 
within the phonon spectrum or within the anti-phonon spectrum do not give rise to 
dynamical instability; they only generate gaps in the spectra.
Based on this general conclusion, we consider two special cases, which allow for
simple explanations of the onset of dynamical instability. 

One case is the weak periodic potential limit $v\ll 1$, where we can 
approximate the boundary with the free space case.
This case corresponds to  the first column of 
Fig. \ref{fig:kq}. In this limit, we can approximate the phonon spectrum 
and the anti-phonon spectrum with the ones given in Eq. (\ref{eq:epsilon}).
By equating them, $\epsilon_{+}(q-1)=\epsilon_{-}(q)$, for the degeneracy,
we find that the dynamical instability should occur on the following curves 
\begin{equation} \label{eq:kqc}
k=\sqrt{q^2 c+q^4/4}+\sqrt{(q-1)^2 c+(q-1)^4/4}\,.
\end{equation}
These curves are plotted as solid dots in Fig.\,\ref{fig:kq},
and they fall right in the middle of the thin dark-shaded areas. To
some extent, one can regard these thin dark-shaded areas as broadening
of the curves (\ref{eq:kqc}). It is noted in Ref. ~\cite{smith2} that
the relation (\ref{eq:kqc}) is also the result of
$\epsilon_+(q-1)+\epsilon_+(-q)=0$,
which involves only the physical phonons. Therefore, the physical meaning
of Eq. (\ref{eq:kqc}) is that one can excite a pair of phonons with total energy 
zero and with total momentum equal to a reciprocal wave number of the lattice.

The other case, $c\ll v$, is shown in the first row of 
Fig. \,\ref{fig:kq}. The open circles along
the left edges of these dark-shaded areas are given by
\begin{equation}\label{eq:bb}
E_1(k+q)-E_1(k)=E_1(k)-E_1(k-q),
\end{equation}
where  $E_1(k)$ is the lowest Bloch band of 
\begin{equation}
H_0=-{1\over 2}{\partial^2\over \partial x^2}+v\cos (x).
\end{equation}
In this linear periodic system, the excitation spectrum (phonon or anti-phonon) 
just corresponds to transitions from the Bloch states of energy $E_1(k)$
to other Bloch states of energy $E_n(k+q)$,  or vise versa.  
The above equation is just the resonance condition between such excitations in
the lowest band ($n=1$).  Alternatively, we can write the resonance 
condition as
\begin{equation}
E_1(k)+E_1(k)=E_1(k+q)+E_1(k-q)\,.
\end{equation}
So, this condition may be viewed as the energy and momentum conservation 
for two particles interacting and decaying into two 
different Bloch states $E_1(k+q)$ and $E_1(k-q)$.  This is
 the same physical picture behind Eq. (\ref{eq:kqc}).

One common feature of all the diagrams in Fig.\,\ref{fig:kq} is that
there are two  critical Bloch wave numbers, $k_t$ and $k_d$. 
Beyond $k_t$ the Bloch waves $\psi_k$ suffer the Landau instability; 
beyond $k_d$ the Bloch waves $\psi_k$ are dynamically unstable.
The onset of instability at $k_d$ always corresponds to $q=1/2$.
In other words, if we drive the Bloch state $\psi_k$ from $k=0$ to $k=1/2$
the first unstable mode appearing is always $q=\pm 1/2$, which 
represents period doubling.  Only for $k>k_d$ can longer wavelength
instabilities occur. The growth of these unstable modes drives the system 
far away from the Bloch state and spontaneously breaks the translational
symmetry of the system.  

\subsection{Two critical velocities}
\begin{figure}
\begin{center}
\includegraphics[width=8cm]{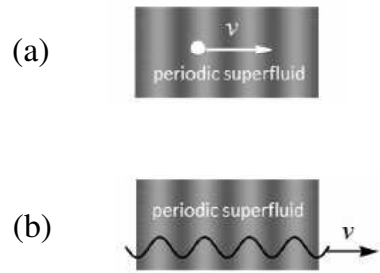}
\caption{(a) A macroscopic obstacle moves with a  velocity of ${\bf v}$
inside a superfluid residing in a periodic potential. The superfluid and periodic potential
are ``locked'' together and there is no relative motion between them.
(b) The lattice where a superfluid resides is slowly accelerated 
to a velocity of ${\bf v}$. }
\label{fig:slatt}
\end{center}
\end{figure}

Besides inducing the dynamical instability, the presence of the optical lattice has also
non-trivial consequences on the concept
of  critical velocity.  In contrast to the homogeneous
superfluid which has only one critical velocity, there are two
distinct critical velocities for a periodic superfluid~\cite{twocv}.
The first one, which we call inside critical velocity, is for an
impurity to move frictionlessly in the periodic superfluid
system (Fig. \ref{fig:slatt}(a)); the second, which is called trawler
critical velocity, is the largest velocity of the lattice 
for the superfluidity to maintain (Fig. \ref{fig:slatt}(b)).  
We illustrate these two  critical velocities with
a BEC in a one-dimensional optical lattice.

The presence of the optical lattice plays a decisive role in the
appearance of the two critical velocities: two very different situations 
can arise. The first situation is described in
Fig. \ref{fig:slatt}(a), where one macroscopic impurity moves inside
the superfluid. The key feature in this situation is that there is no
relative motion between the superfluid and the lattice. 
The other situation is illustrated in Fig. \ref{fig:slatt}(b),  where the
lattice is slowly accelerated to a given velocity and there is a
relative motion between the superfluid and the lattice.
For these two different situations, naturally arise two critical
velocities.

In the first scenario, we consider a moving impurity that generates an excitation
with momentum ${\bf q}$ and energy $\epsilon_0({\mathbf{q}})$ in the BEC.
According to the conservations of both momentum and energy, we should have
\begin{eqnarray}  \label{con1}
m_0{\mathbf{v}}_i&=&m_0{\mathbf{v}}_f+{\mathbf{q}}+n\hbar{\bf G}\,, \\
\frac{m_0{\mathbf{v}}_i^2}{2}&=&\frac{m_0{\mathbf{v}}_f^2}{2}+\epsilon_0({%
\mathbf{q}})\,,  \label{con2}
\end{eqnarray}
where $m_0$ is the mass of the impurity, ${\mathbf{v}}_i$ and ${\mathbf{v}}_f
$ are the initial and final velocities of the impurity, respectively, 
${\bf G}$ is the reciprocal vector, and $%
\epsilon_0({\mathbf{q}})$ is the excitation of the BEC at the lowest Bloch state $k=0$. 
Note that in contrast to the conservation of momentum in free space, here the total momentum
of the impurity and the excitation is not exactly conserved due to the presence of an optical lattice,
and the momenta differing by integer multiples of reciprocal lattice vector are equivalent.
For phonon excitations, i.e., $\epsilon_0({\mathbf{q}})=c|{%
\mathbf{q}}|$, these two conservations can always be satisfied simultaneously
no matter how small the velocity of the impurity is. In other words, the critical
velocity for this scenario is exactly zero. 

In the other scenario, there is relative motion between the superfluid and the optical lattice,
and the superfluid no longer resides in the $k=0$ point of the Brillouin zone.
We should examine the stability of Bloch waves with nonzero $k$, which is discussed in
detail in the previous subsection. The critical wave number $k_t$ mentioned above 
corresponds  precisely to the trawler critical velocity $v_t$ here. As $k_t$ is not zero, 
$v_t$ is not zero.  

Both critical velocities can be measured with BECs in
optical lattices. The inside critical velocity $v_i$ can be measured
with the same experimental setting as in Ref. ~\cite{Onofrio2000PRL},
where the superfluidity of a BEC was studied by moving a blue-detuned
laser inside the BEC.  For the trawler critical velocity $v_t$, one
can repeat the experiment in Ref. ~\cite{Fallani2004PRL}, where a BEC is
loaded in a moving optical lattice. One only needs to shift his
attention from dynamical instability to superfluidity. The potential difficulty lies
in that the Landau instability occurs over a much larger time scale, which may
be beyond the life time of a BEC~\cite{Fallani2004PRL}. 

\section{Superfluidity with spin orbit coupling}\label{socsuperfluid}

The intrinsic spin-orbit coupling (SOC) of electrons plays a crucial role in many 
exotic materials, such as topological insulators~\cite{TI}. 
In spintronics~\cite{spintronics},  its presence 
enables us to manipulate the spin of electrons by means of exerting 
electric field instead of magnetic field, which is much
easier to implement for industrial applications. However, as a relativistic effect,
the intrinsic SOC  does not exist or is very weak for bosons 
in nature. With the method of engineering atom-laser interaction,
an artificial SOC has been  realized for ultracold
bosonic gases in~\cite{lin2,fu,pan,chen}. A great deal of effort has
been devoted to study  many interesting properties of
spin-orbit coupled  BECs ~\cite%
{shem,zhai1,ho,victor,zhang,yip,zhai2,baym,pu,you,zhangqi,santos,fleischhauer}. 

\begin{figure}[t]
\includegraphics[width=8cm]{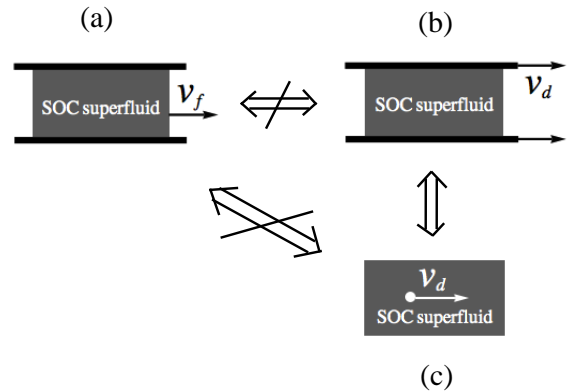}\newline
\caption{(a) A superfluid with SOC moves while the tube is
at rest. (b) The superfluid is dragged by a tube moving at the speed of $v$.
(c) An impurity moves at $v$ in the SOC superfluid. The reference frame is
the lab. The two-way arrow indicates the equivalence between different
scenarios and the arrow with a bar indicates the non-equivalence.}
\label{soc}
\end{figure}

A dramatic change that the SOC brings to the concept of 
superfluidity is the breakdown of Landau's criterion of critical velocity (\ref{min}) 
and the appearance of two different critical velocities. Laudau's criterion 
of critical velocity (\ref{min}) is based on the Galilean invariance.
It is apparent to many that the scenario where a superfluid flows inside a motionless 
tube is equivalent to the other scenario where a superfluid at rest is
dragged by a moving tube.  If the flowing superfluid loses its superfluidity 
when its speed exceeds a critical speed $v_c$, then the superfluid 
in the other scenario will be dragged into motion by a tube moving 
faster than $v_c$.  However, this equivalence is based on
that the superfluid is invariant under the Galilean transformation. 
As SOC breaks the Galilean invariance of the system~\cite{messiah}, 
we find that the two scenarios mentioned above are no longer equivalent 
as shown in Fig. \ref{soc}: the critical speed for scenario (a) is different 
from the one for scenario (b). For easy reference, the critical speed for (a) 
is hereafter called the critical flowing speed and the one for (b) the 
critical dragging speed.

For ultra-cold atomic gases, the breakdown of the Galilean invariance at the 
presence of  SOC  can be understood both theoretically and experimentally. 

From the theoretical point of view,
we show in detail how a system with SOC changes under the Galilean transformation.
We adopt the formalism in Ref. ~\cite{messiah}. The operator for the Galilean transformation is
\begin{equation}
G(\textbf{v},t)=\exp\left[i\textbf{v}\cdot(m\textbf{r}-\textbf{p}t)/\hbar\right],
\end{equation}
which satisfies the definition
\begin{eqnarray}
G^{\dagger}(\textbf{v},t)\,\textbf{r}\,G(\textbf{v},t)=&\textbf{r}-\textbf{v}t\,,\\
G^{\dagger}(\textbf{v},t)\,\textbf{p}\,G(\textbf{v},t)=&\textbf{p}-m\textbf{v}\,,\\
G^{\dagger}(\textbf{v},t)\bm{\sigma}G(\textbf{v},t)=&\bm{\sigma}\,.
\end{eqnarray}
A system is invariant under the Galilean transformation if the following equation is satisfied (Ref. ~\cite{messiah}),
\begin{equation}\label{galilean}
   G^{\dagger}(\textbf{v},t)\left[i\hbar\frac{\partial}{\partial t}-H\right]G(\textbf{v},t)
  =\left[i\hbar\frac{\partial}{\partial t}-H\right]\,.
\end{equation}
The above condition is clearly satisfied by a Hamiltonian without SOC, e.g., $H=\textbf{p}^2/2m$.
However, for a Hamiltonian with SOC, e.g., $H_{\textrm{soc}}=\textbf{p}^2/2m+\gamma\left(\sigma_x p_y-\sigma_y p_x\right)$, it is easy to check that 
   \begin{equation}\label{galilean}
   G^{\dagger}(\textbf{v},t)\left[i\hbar\frac{\partial}{\partial t}-H_{\textrm{soc}}\right]G(\textbf{v},t)
  =\left[i\hbar\frac{\partial}{\partial t}-H_{\rm soc}'\right]\,,
\end{equation} 
where $H_{\rm soc}'=H_{\textrm{soc}}+m\gamma\left(\sigma_xv_y-\sigma_yv_x\right)$. Clearly there is an additional
term dependent on the velocity of the reference frame. This new term can be regarded as an
effective Zeeman effect and can not be gauged away; the  
 Galilean invariance of the system is thus lost.

In the experiments of ultra-cold atomic gases, the SOC is created by two 
Raman beams that couple two hyperfine states of the atom. Since the Galilean transformation only boosts the BEC, not including the laser setup
as a whole, the moving BEC will experience a different laser field due to the Doppler effect, 
resulting a loss of  the Galilean invariance.

\subsection{Bogoliubov excitations and definition of critical velocities}
We use the method introduced in Sec. \ref{gp} to study the superfluidity 
of a BEC with SOC by computing its elementary excitations \cite{socepl}. 
In experiments, only the equal combination of Rashba and Dresselhaus coupling is realized.
Here we use the Rashba coupling as an example as the main conclusion 
does not rely on the details of the SOC type. 

We calculate how the elementary excitations change with the flow speed and
manage to derive from these excitations the critical speeds for the two
different scenarios shown in Fig. \ref{soc}(a,b). We find that there are two
branches of elementary excitations for a BEC with SOC: the lower branch is
phonon-like at long wavelengths and the upper branch is generally gapped.
Careful analysis of these excitations indicates that the critical flowing velocity 
for a BEC with SOC (Fig. \ref{soc}(a)) is non-zero 
while the critical dragging speed is zero (Fig. \ref{soc}(b)). 
This shows that the critical velocity
depends on the reference frame for a BEC with SOC and, probably, for
any superfluid that has no Galilean invariance.

Specifically, we consider a BEC with pseudo-spin $1/2$ and the Rashba SOC.
The  system can be described by the Hamiltonian ~\cite{zhai1,merkl,larson,zhang}
\begin{eqnarray}
\mathcal{H} &=&\int d\mathbf{r}\left\{\sum_{\sigma =1,2}\psi _{\sigma
}^{\ast }\left(-\frac{\hbar^{2}\nabla^{2}}{2M}+V(\mathbf{r})\right)\psi
_{\sigma}\right.   \notag \\
&&\left. +\gamma\left[\psi_{1}^{\ast }(i\hat{p}_{x}+\hat{p}_{y})\psi
_{2}+\psi_{2}^{\ast}(-i\hat{p}_{x}+\hat{p}_{y})\psi_{1}\right]\right.
\notag \\
&&\left. +\frac{C_{1}}{2}\left(|\psi_{1}|^{4}+|\psi_{2}|^{4}\right)
+C_{2}|\psi_{1}|^{2}|\psi_{2}|^{2}\right\}\,,  \label{ham}
\end{eqnarray}%
where $\gamma$ is the SOC constant, $C_{1}$ and $C_{2}$ are interaction
strengths between the same and different pseudo-spin states, respectively.
For simplicity and easy comparison with previous theory, we 
focus on the homogeneous case $V(\mathbf{r})=0$ despite that the BEC
usually resides in a harmonic trap in experiments. Besides, we limit ourselves
to the case $C_{1}>C_{2}$, namely, in the plane wave phase.
In the following discussion, for simplicity, we set $%
\hbar=M=1$ and ignore the non-essential $z$ direction, treating the system
as two-dimensional. We also assume the BEC moves in the $y$ direction, and the
critical velocity is found to be not influenced by the excitation in the $z$ direction.

\begin{figure*}[!htb]
\center{\includegraphics[width=14cm]{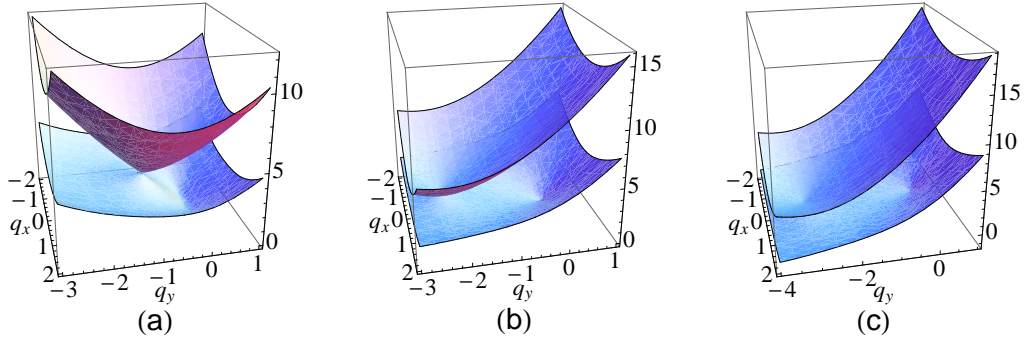}}
\caption{(color online)Elementary excitations of a BEC flow with the SOC.
(a) $k=1$; (b) $k=3$; (c) $k=4$. $C_1=10$, $C_2=4$, $\protect\gamma=1$. }
\label{3D}
\end{figure*}

The GP equation obtained from the Hamiltonian (\ref{ham}) has
plane wave solutions
\begin{equation}
\phi_{{\mathbf{k}}}=\left(%
\begin{array}{c}
\psi_1 \\
\psi_2 %
\end{array}%
\right)
=\frac{1}{\sqrt{2}}\left(%
\begin{array}{c}
e^{i\theta_{\mathbf{k}}} \\
- 1%
\end{array}%
\right)e^{i\mathbf{k}\cdot\mathbf{r}-i\mu(\mathbf{k}) t}\,,  \label{pwave}
\end{equation}
where $\tan\theta_{\mathbf{k}}=k_{x}/k_{y}$, $\mu(\mathbf{k})=|\mathbf{k}%
|^2/2-\gamma|\mathbf{k}|+(C_1+C_2)/2$. The solution $\phi_{{\mathbf{k}}}$ is
the ground state of the system when $|\mathbf{k}|=\gamma$. There are another
set of plane wave solutions, which have higher energies and are not relevant
to our discussion.

We study first the scenario depicted in Fig. \ref{soc}(a), 
where the BEC flows with a given velocity. Since the system is
not invariant under the Galilean transformation, we cannot use Laudau's argument
 to find the excitations for the flowing BEC from the excitation of
a stationary BEC. We have to compute the excitations directly. This can be
done by computing the elementary excitations of the state $\phi _{{\mathbf{k}}}$
 with the Bogoliubov equation for different values of $\mathbf{k}$.

Without loss of generality, we choose $\mathbf{k}=k\hat{y}$ with $k>0$.
Following the standard procedure of linearizing the GP
equation ~\cite{wu1,wu2}, we have the following Bogoliubov equation%
\begin{equation}
\mathcal{M}\left(
\begin{array}{c}
u_{1} \\
u_{2} \\
v_{1} \\
v_{2}%
\end{array}%
\right) =\epsilon \left(
\begin{array}{c}
u_{1} \\
u_{2} \\
v_{1} \\
v_{2}%
\end{array}%
\right) ,
\end{equation}%
where
\begin{equation}
\mathcal{M}=\left(
\begin{array}{cccc}
H_{k}^{+} & b_{12} & -\frac{1}{2}C_{1} & -\frac{1}{2}%
C_{2} \\
 b_{21} & H_{k}^{+} & -\frac{1}{2}C_{2} & -\frac{1}{2}%
C_{1} \\
\frac{1}{2}C_{1} & \frac{1}{2}C_{2} & H_{k}^{-} &  b_{34} \\
\frac{1}{2}C_{2} & \frac{1}{2}C_{1} & b_{43} &
H_{k}^{-}%
\end{array}%
\right),
\end{equation}%
with $H_{k}^{\pm }=\pm \frac{q_{x}^{2}+(q_{y}\pm k)^{2}}{2}\pm A$, $A=\frac{%
C_{1}}{2}-\frac{k^{2}}{2}+\gamma k$, $b_{12}=-\gamma(iq_{x}+q_{y}+k)+\frac{C_{2}}{2}$, $%
b_{21}=\gamma(iq_{x}-q_{y}-k)+\frac{C_{2}}{2}$,
 $b_{34}=\gamma(iq_{x}-q_{y}+k)-\frac{C_{2}}{2}$, and $b_{43}=-\gamma(iq_{x}+q_{y}-k)-\frac{C_{2}}{2}$.
 As usual, there are two groups of eigenvalues and only the ones whose
corresponding eigenvectors satisfy $|u_{i}|^{2}-|v_{i}|^{2}=1$ ($i=1,2$) are
physical.

\begin{figure}[!htb]
\includegraphics[width=3.4in]{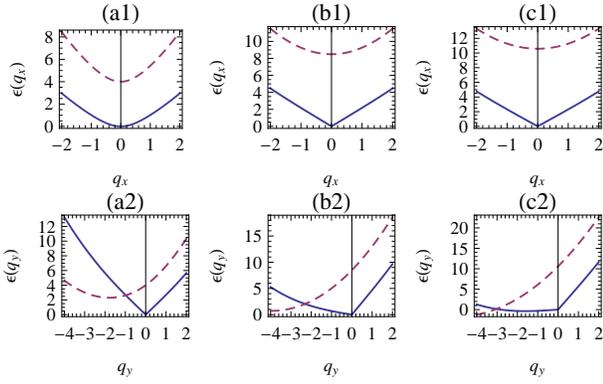}\newline
\caption{Excitations along the $x$ axis (the first row) and $y$
axis (the second row) at different values of $k$. (a1,a2) $k=1$; (b1,b2)
 $k=3$; (c1,c2) $k=4$. $C_1=10$, $C_2=4$, $\protect\gamma=1$.}
\label{xy}
\end{figure}

In general there are no simple analytical results. We have numerically
diagonalized $\mathcal{M}$ to obtain the elementary excitations. We find
that part of the excitations are imaginary for BEC flows with $|{\mathbf{k}}%
|<\gamma$. This means that all the flows with $|{\mathbf{k}}|<\gamma$ are
dynamically unstable and therefore do not have superfluidity. For other
flows with $k\ge \gamma$, the excitations are always real and they are
plotted in Fig. \ref{3D}. One immediately notices that the excitations have
two branches, which contact each other at a single point. Closer
examination shows that the upper branch is gapped in most of the cases while
the lower branch has phonon-like spectrum at large wavelength. These
features are more apparent in Fig. \ref{xy}, where only the excitations
along the $x$ axis and $y$ axis are plotted.

\begin{figure}[!t]
\includegraphics[width=3.4in]{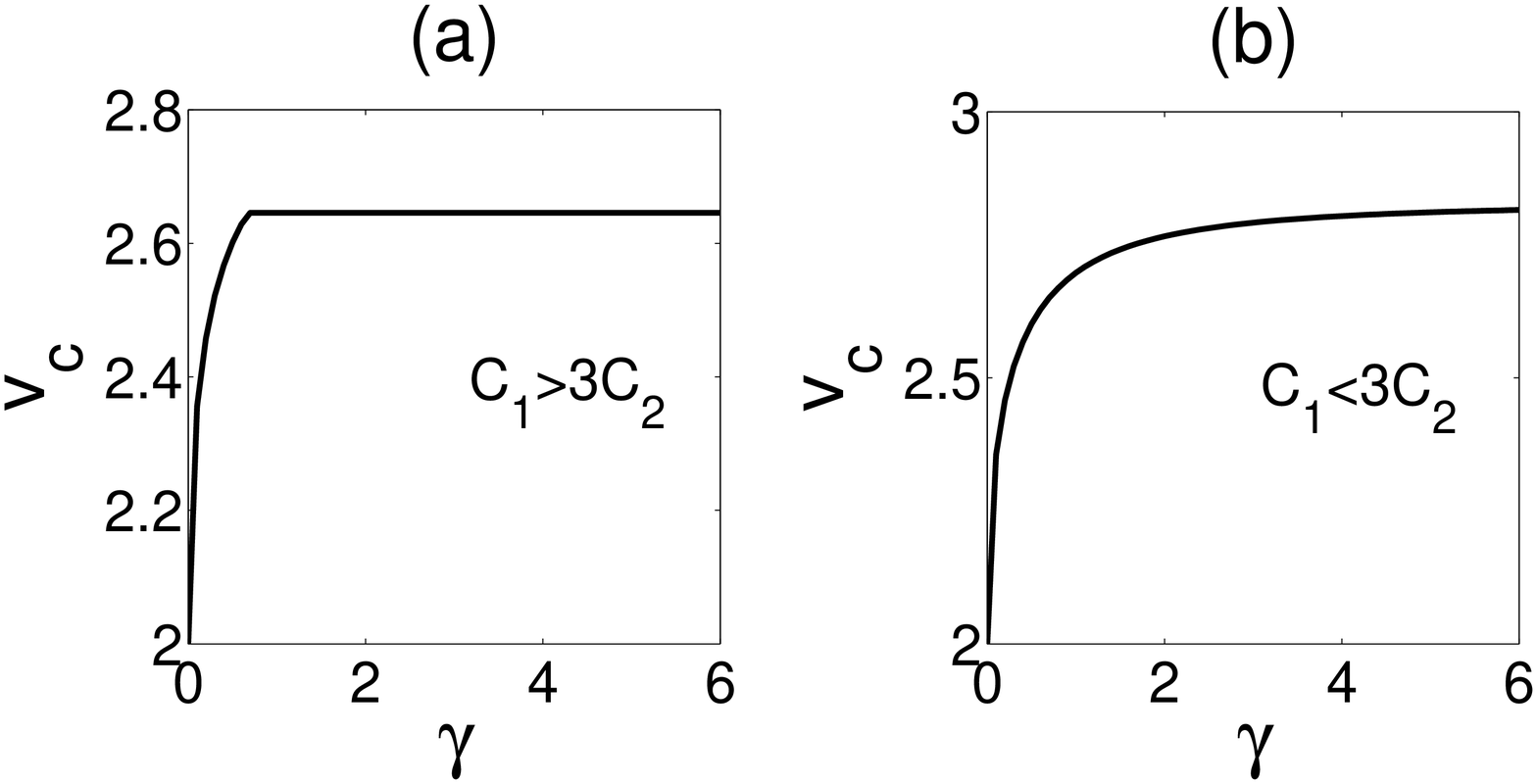}\newline
\caption{Critical flowing velocity $v_c$ of a BEC as a function
of the SOC parameter $\protect\gamma$. (a) $C_1=11$, $C_2=3$; (b) $C_1=14$, $C_2=6$.}
\label{vc}
\end{figure}

In Fig. \ref{3D}(c) and Fig. \ref{xy}(c2), we notice that part of the
excitations in the upper branch are negative, indicating that the underlying
BEC flow is thermodynamically unstable and has no superfluidity. In fact,
our numerical computation shows that there exists a critical value $k_{c}$:
when $k>k_{c}$ either part of the upper branch of excitations or part of the
lower branch or both become negative. This means that the flows described by
the plane wave solution $\phi _{{\mathbf{k}},-}$ with $|{\mathbf{k}}|>k_{c}$
suffer Landau instability and have no superfluidity. Combined with the
fact that the flows with $|{\mathbf{k}}|<\gamma $ are dynamically unstable,
we can conclude that only the flows with $\gamma \leq \mathbf{|k|}\leq k_{c}$
have superfluidity. The corresponding critical speed is $v_{c}=k_{c}-\gamma $. 
We have plotted how the critical flowing velocity $v_{c}$ varies
with the SOC parameter $\gamma$ in Fig. \ref{vc}.

We turn to another reference frame illustrated in Fig. \ref{soc}(b), where
the BEC can be viewed as being dragged by a moving tube. To simplify the
discussion, we replace the moving tube with a macroscopic impurity moving
inside the BEC as shown in Fig. \ref{soc}(c). Correspondingly, the question
``whether the BEC will be dragged along by the moving tube?" is replaced by
an equivalent question ``whether the impurity will experience any
viscosity?". Suppose that the moving impurity generates an excitation in the BEC.
According to the conservations of both momentum and energy, we should have
\begin{eqnarray}  \label{con1}
m_0{\mathbf{v}}_i&=&m_0{\mathbf{v}}_f+{\mathbf{q}}\,, \\
\frac{m_0{\mathbf{v}}_i^2}{2}&=&\frac{m_0{\mathbf{v}}_f^2}{2}+\epsilon_0({%
\mathbf{q}})\,,  \label{con2}
\end{eqnarray}
as those for a BEC without SOC, where $%
\epsilon_0({\mathbf{q}})$ is the excitation of the BEC at $k=\gamma$. 
The critical dragging velocity derived from Eqs. (\ref{con1},\ref{con2}) is given by
\begin{equation}
v_c=\left|\frac{\epsilon_0({\mathbf{q}})}{|{\mathbf{q}}|}\right|_{min}\,.
\label{cerenkov}
\end{equation}
If the excitations were purely phonons, i.e., $\epsilon_0({\mathbf{q}})=c|{%
\mathbf{q}}|$, these two conservations would not be satisfied simultaneously
when $v\approx|{\mathbf{v}}_i|\approx|{\mathbf{v}}_f|<c$. This means that
the impurity could not generate phonons in the superfluid and would not
experience any viscosity when its speed was smaller than the sound speed.
Unfortunately,  for our BEC system, the elementary excitations $\epsilon_0({%
\mathbf{q}})$ are not purely phonons, as will be shown below.

When $\gamma\neq 0$, the excitations $\epsilon_0({\mathbf{q}})$ also
share two branches. Along the $x$ axis, these two branches are
\begin{equation}
\epsilon _{0}^{\pm }(q_{x})=\sqrt{s_{1}+s_{2}q_{x}^{2}+\frac{q_{x}^{4}}{4}%
\pm \sqrt{t_{1}+t_{2}q_{x}^{2}+t_{3}q_{x}^{4}+\gamma ^{2}q_{x}^{6}}},
\end{equation}%
where $s_{1}=2\gamma ^{4}+\gamma ^{2}\left( C_{1}-C_{2}\right) $, $%
s_{2}=2\gamma ^{2}+\frac{1}{2}C_{1}$, $t_{1}=s_{1}^{2}$, $t_{2}=2s_{1}s_{2}$%
, and $t_{3}=2s_{1}+\left( \gamma ^{2}+C_{2}/2\right) ^{2}$.
Along the $y$ axis, the excitations of the ground state are
\begin{eqnarray}
\epsilon _0^{-}(q_{y}) &=&\sqrt{\frac{C_{1}+C_{2}}{2}q_{y}^{2}+\frac{q_{y}^{4}%
}{4}}\,, \\
\epsilon _0^{+}(q_{y}) &=&2\gamma q_{y}+\sqrt{2s_{1}+\left( s_{2}-\frac{C_{2}}{%
2}\right) q_{y}^{2}+\frac{q_{y}^{4}}{4}}\,.
\end{eqnarray}%
When $\gamma >0$, the upper branch $\epsilon _{0}^{+}(q_{x})$ is always parabolic at
small $q_{x} $ with a gap $\sqrt{2s_{1}}$. When expanded to the second order
of $q_{x}$, the lower branch has the following form
\begin{equation}
\epsilon _{0}^{-}(q_{x})\approx q_{x}^{2}\sqrt{\frac{C_{1}+C_{2}}{8\gamma
^{2}}}\,.  \label{para}
\end{equation}%
This shows that $\epsilon _{0}^{-}(q_{x})$ is parabolic at long
wavelengths instead of linear as usually expected for a boson system. This
agrees with the results in Ref. ~\cite{zhai2}. This parabolic excitation has
a far-reaching consequence: according to Eq. (\ref{cerenkov}), the critical
dragging velocity $v_c$ is zero, very different from the critical flowing velocity
for a BEC moving in a tube. This shows that the critical velocity for a BEC
with SOC is not independent of the reference frame, in stark contrast with a
homogeneous superfluid without SOC. This result of course has
the root in the fact that the BEC described by the SOC Hamiltonian (\ref{ham}%
) is not invariant under the Galilean transformation ~\cite{messiah}.

We have also investigated the superfluidity with the
general form of SOC, which is a mixture of Rashba and Dresselhaus coupling.
Mathematically, this SOC term has the form $\alpha\sigma_{x}p_{y}-\beta
\sigma_{y}p_{x}$. The essential physics is the same: the critical flowing
speed is different from the critical dragging speed, and therefore the
critical velocity depends on the choice of the reference frame. However, the
details do differ when $\alpha \neq \beta $. The critical dragging
speed is no longer zero.  Without loss of generality, we let $\alpha >\beta$. The slope
of the excitation spectrum for the ground state  along the $y$ axis is 
\begin{equation}
v_{y}=\sqrt{\sqrt{2\alpha^{2}\left(C_{1}-C_{2}+2\alpha^{2}\right)}+2\alpha^{2}+\frac{C_{1}-C_{2}}{2}}-2\alpha\,, 
\end{equation}
and the slope along  the $x$ axis,
\begin{equation}
v_{x}=\sqrt{\left(1-\frac{\beta^2}{\alpha^2}\right)\frac{C_1+C_2}{2}}\,.
\end{equation} 
The critical dragging velocity is  the smaller one of the above two slopes, both 
of which are nonzero.

\subsection{Experimental observation}
Spin-orbit coupled BECs
have been realized by many different groups~\cite{lin2,fu,pan,chen}
through coupling ultracold $^{87}$Rb~atoms with laser fields. The strength
of the SOC in the experiments can be tuned by changing the directions of the
lasers~\cite{lin2,fu,pan} or through the fast modulation of the laser
intensities~\cite{yongping2}. The interaction between atoms can be adjusted
by varying the confinement potential, the atom number or through the
Feshbach resonance~\cite{feshbach}. For the scenario in Fig. \ref{soc}(b), one
can use a blue-detuned laser to mimic the impurity  for the measurement of the
critical dragging speed similar to the experiment  in Ref.~\cite{raman}.
For the scenario in Fig. \ref{soc}(a), there are two possible experimental setups for
measuring the critical flowing speed.  In the first one, one generates
a dipole oscillation similar to the experiment in Ref. ~\cite{pan} but with a blue
detuned laser inserted in the middle of the trap. The second one is more complicated:
At first, one can generate a moving BEC with a gravitomagnetic
trap~\cite{motion}. One then uses Bragg spectroscopy ~\cite%
{brag1,brag2} to measure the excitations of the moving BEC, from which the
superfluidity can be inferred. For the typical atomic density
of $10^{14}\sim10^{15}$ cm$^{-3}$ achievable in current experiments~\cite{raman}, 
and the experimental setup in Ref.~\cite{lin2},
the critical flowing velocity is $0.2\sim0.6$ mm/s, while the critical dragging 
velocity is still very small, about $10^{-3}\sim10^{-2}$ mm/s. 
To further magnify the difference between the two critical
velocities, one may use the Feshbach resonance to tune the $s$-wave 
scattering length.

\section{Superfluidity of spin current}\label{spincurrent} 
A  neutral boson can carry both mass and spin; it thus can carry both
mass current and spin current. However, when a boson 
system is said to be a superfluid, it traditionally refers only to its mass current. 
The historical reason is that the first superfluid discovered in experiment
is the spinless helium 4 which carries only mass current. 
For a boson with spin, say, a spin-1 boson, 
we can in fact have a pure spin current, a spin current with no mass current.  
This pure spin current can be generated by putting an unpolarized 
spin-1 boson system in a magnetic field with a small gradient. 
{\it Can such a pure spin current be a super-flow?} In this section
we try to address this issue for both planar and circular pure spin currents
by focusing on an unpolarized spin-1 Bose gas. 

The stability of a pure spin current in an unpolarized spin-1 Bose gas was 
studied in~\cite{fujimoto}.  It was found that such a current is generally unstable
and is not a super-flow. We have recently found that the pure spin current 
can be stabilized to become a super-flow~\cite{unpub}: ({\it i}) for a planar flow,
it can be stabilized by quadratic Zeeman effect;  ({\it ii}) for a circular flow, it can be
stabilized with SOC. We shall discuss these results and related 
experimental schemes in detail in the next subsections. 

There has been lots of study on the counterflow in a two-species 
BEC~\cite{law,kuklov,yukalov,takeuchi,engels,hamner,fujimoto,ishino,jltp,abad,kurn,cherng}, which appears very similar
to the pure spin current discussed here.  It was found that 
there is a critical relative speed  beyond which the counterflow state loses
its superfluidity and becomes unstable~\cite{law,kuklov,yukalov,takeuchi,engels,hamner,ishino,jltp,abad}. 
We emphasize that the counterflow in a two-species 
BEC is not a pure spin current for two reasons.  Firstly, although the two species 
may be regarded as two components of a psudo-spin 1/2,  they do not have SU(2) 
symmetry. Secondly, it is hard to prepare experimentally a BEC with exactly equal
numbers of bosons in the two species to create  a counterflow with no mass current.
There is also interesting work addressing the issue of spin superfluidity in other situations~\cite{egor,flayac,weibin}.

\subsection{Planar flow}\label{planar}
We consider a spin-1 BEC in free space. The mean-field wave function of 
such a spin-1 BEC satisfies the following GP equation \cite{review},
\begin{equation}
\label{ }
i\hbar\frac{\partial}{\partial t}\psi_m=-\frac{\hbar^2\nabla^2}{2M}\psi_m+c_0 \rho\psi_m+c_2\sum_{n=-1}
^1\mathbf{s}\cdot\mathbf{S}_{mn}\psi_n,
\end{equation}
where $\psi_m$ ($m=1,0,-1$) are the components of the macroscopic wave function.
$\rho=\sum_{m=-1}^1|\psi_m|^2$ is the total density, ${\bf s}_i=\sum_{mn}\psi_m^*(S_i)_{mn}\psi_n$
is the spin density vector and ${\bf S}=(S_x, S_y, S_z)$ is the spin operator vector with $S_i$ ($i=x,y,z$)
being the three Pauli matrices in the spin-1 representation.
The collisional interactions include a spin-independent
part $c_{0}=4\pi\hbar^{2}(a_{0}+2a_{2})/3M$ and a spin-dependent
part $c_{2}=4\pi\hbar^{2}(a_{2}-a_{0})/3M$, with $a_{f}(f=0,2)$
being the $s$-wave scattering length for spin-1 atoms in the
symmetric channel of total spin $f$.

We consider a spin current state of the above GP equation with the form
\begin{equation}
\label{scurrent}
\psi=\sqrt{\frac{n}{2}}\left(
\begin{array}{c}
e^{i\mathbf{k}_1\cdot{\bf r}}\\  0\\
e^{i\mathbf{k}_2\cdot{\bf r}}
\end{array}
\right)\,,
\end{equation}
where $n$ is the density of the uniform BEC.
The requirement of equal chemical potential leads to $|\mathbf{k}_1|=|\mathbf{k}_2|$.
In the case where $\mathbf{k}_1=-\mathbf{k}_2$,  this state carries a pure spin current:
the total mass current is zero as  it has equal mass counterflow while the spin current is nonzero.

To determine whether the state (\ref{scurrent}) represents a superfluid, we need compute its
Bogoliubov excitation spectrum, also using the method introduced in Sec. \ref{gp}.
It is instructive to first consider the special case when there is no counterflow, i.e., ${\bf k}_1={\bf k}_2=0$.
The excitation spectra are found to be $\epsilon^0=\sqrt{2c_2n\epsilon_q+\epsilon_q^2}$ and $\epsilon_1^{\pm1}=\sqrt{2c_0n\epsilon_q+\epsilon_q^2}$, $\epsilon_2^{\pm1}=\sqrt{2c_2n\epsilon_q+\epsilon_q^2}$, respectively,
with $\epsilon_q=\hbar^2q^2/2M$.
So for antiferromagnetic interaction ($c_0>0, c_2>0$), all branches of the spectra are real
and  there is a double degeneracy in one branch of the spectra. The phonon excitations give two sound velocities, $\sqrt{n c_i/M}$ ($i=0,2$), corresponding to the speeds of density wave and spin wave, respectively. However, the existence of phonon excitation
does not mean that  the pure spin current (${\bf k}_1={\bf k}_2\neq 0$) is a super-flow
as we can not obtain the current with ${\bf k}_1={\bf k}_2\neq 0$ from the state
with ${\bf k}_1={\bf k}_2= 0$ by a Galilean transformation.

For the counterflow state with $\mathbf{k}_1=-\mathbf{k}_2\neq 0$,
the stability has been studied in Ref. \cite{fujimoto}. 
It is found that, for  the antiferromagnetic interaction case ($c_0>0, c_2>0$), the excitation
spectrum of the $m=0$ component always has nonzero imaginary part
in the long wavelength limit as long as there is counterflow between the two components, and the imaginary excitations in the $m=1,-1$ components only appear for a large enough relative velocity $v_1=2\sqrt{n c_2/M}$. For the ferromagnetic
interaction case ($c_0>0, c_2<0$), both excitation spectra of the $m=0$ and $m=1,-1$ components have nonzero imaginary parts for any relative velocity. This means that the pure spin current cannot be stable in any cases.

For the general non-collinear case (${\bf k}=\frac{{\bf k}_1+{\bf k}_2}{2}\neq0$) and antiferromagnetic interaction, the excitation spectrum for the $m=0$ component is found to be
\begin{equation}
\label{ }
\epsilon^0=\sqrt{\left(\epsilon_q+\frac{\hbar^2}{2M}\left(|\mathbf{k}|^2-|\mathbf{k}_1|^2\right)+c_2n\right)^2-c_2^2n^2}.
\end{equation}
We see here that as long as the momenta of the two components are not exactly parallel, i.e., $\mathbf{k}_1$ is not exactly equal to $\mathbf{k}_2$, then $|\mathbf{k}|<|\mathbf{k}_1|$, and
there is always dynamical instability for the long wavelength excitations.

Therefore, the spin current in Eq. (\ref{scurrent}) is generally unstable and not a super-flow.
This instability  originates from the interaction process described by
$\psi_0^{\dagger}\psi_0^{\dagger}\psi_1\psi_{-1}$ in the second
quantized Hamiltonian. This energetically favored process converts two
particles in the $m=1,-1$ components, respectively,  into two stationary particles
in the $m=0$ component. To suppress such a process and achieve a stable
pure spin current, one can utilize the quadratic Zeeman effect.
With the quadratic Zeeman effect of negative coefficient, the Hamiltonian
adopts an additional term $\lambda m^2$ ($\lambda<0$ and $m=1,0,-1$).
This term does not change the energy of the $m=0$
component, but lowers the energy of the other two components $m=1,-1$.
As a result, there arises a barrier for two atoms in the $m=1,-1$ components scattering to the $m=0$ component, and the scattering process is thus suppressed.

The above intuitive argument can be made more rigorous and quantitative.
Consider the case $\mathbf{k}_1=-\mathbf{k}_2$. With the quadratic Zeeman term,
the  excitation spectrum for the $m=0$ component changes to
\begin{equation}
\label{ }
\epsilon^0=\sqrt{\left(\epsilon_q-\frac{\hbar^2 |\mathbf{k}_1|^2}{2M}+c_2n-\lambda\right)^2-c_2^2n^2}.
\end{equation}
So as long as $-\lambda-\hbar^2 |\mathbf{k}_1|^2/2M>0$, long wavelength excitations will be stable for the $m=0$ component. From the excitation
of the $m=0$ component, one can obtain a critical relative velocity of the spin current,
$v_0=2\sqrt{-2\lambda/M}$.  There is  another nonzero critical velocity $v_1=2\sqrt{n c_2/M}$ determined by the
excitations of the $m=1,-1$ components.  The overall critical velocity of the system is the smaller
one of $v_0$ and $v_1$.  Therefore, below the critical relative velocity $v_{\rm c}={\rm min}\{v_0,v_1\}$, the pure spin current is stable and a super-flow. The experimental scheme
to realize such a Zeeman effect will be discussed  in subsection \ref{expe}.

\subsection{Circular flow}\label{circular}
In the cylindrical geometry, we consider a pure spin current formed by two vortices
 with opposite circulation
in the $m=1,-1$ components. From similar arguments, one can expect that interaction will make
such a current unstable.  Inspired by the quadratic Zeeman effect method above,
we propose to use SOC to stabilize it. The SOC can be
viewed as a momentum-dependent effective magnetic field that exerts only on
the $m=1,-1$ components. Therefore, it is possible that SOC lowers
the energy of $m=1,-1$ components, and consequently suppresses the interaction process
leading to the instability.

The model of spin-1 BEC subject to Rashba SOC
can be described by the following energy functional,
\begin{align}
\label{energy}
\mathcal{E}\left[\psi_{\alpha}\right]= & \int d\mathbf{r}\Bigg\{\sum_{\alpha=-1,0,1}\frac{\hbar^{2}|\nabla\psi_{\alpha}|^{2}}{2M}+\rho V(r)+\frac{c_{0}}{2}\rho^{2}
+\frac{c_{2}}{2}\mathbf{s}^{2} \nonumber \\
& +\gamma\langle S_xp_y-S_yp_x\rangle\Bigg\},
\end{align}
where $\rho$ is the density, $V(r)=\frac{1}{2}M\omega^2(x^2+y^2)$ is the trapping potential, and $\gamma$ is the strength of SOC. $\langle\cdots\rangle$ is the expectation value taken with respect to the three component wave function
$\psi=(\psi_1,\psi_0,\psi_{-1})^T$. The SOC strength $\gamma$ defines a characteristic length
$a_{\rm soc}=\hbar/M\gamma$, and can be rescaled to be dimensionless with respect to the harmonic oscillator length
$a_{\rm h}=\sqrt{\hbar/M\omega}$. Then we characterize the strength of SOC with the dimensionless
quantity $\kappa=a_{\rm h}/a_{\rm soc}=\gamma\sqrt{M/\hbar\omega}$.
The SOC of Rashba type here can be generated in various ways, which will be discussed in the next subsection.

The above model can describe a spin-1 BEC of $^{23}$Na confined in a
two-dimensional harmonic trap. Assume the atom number is about $10^6$.
Using the estimate of scattering lengths $a_0=50a_B$, $a_2=55a_B$ \cite{epjd},
with $a_B$ being the Bohr radius, the ground state of spin-1 $^{23}$Na should be
antiferromagnetic because $c_0>0, c_2>0$ \cite{ho}.
Previous studies of spin-1 BEC with Rashba SOC mostly focus on
the strong SOC regime, where the ground
state is found to be the plane wave phase or the stripe phase, for ferromagnetic interaction and antiferromagnetic interaction, respectively \cite{zhai1}.
Here we are interested in the antiferromagnetic interaction case and the
weak SOC regime ($\kappa\ll 1$), and calculate the ground
state wave function of the energy functional with the method of imaginary
time evolution.

\begin{figure}[t]
\begin{center}
\includegraphics[width=8.5cm]{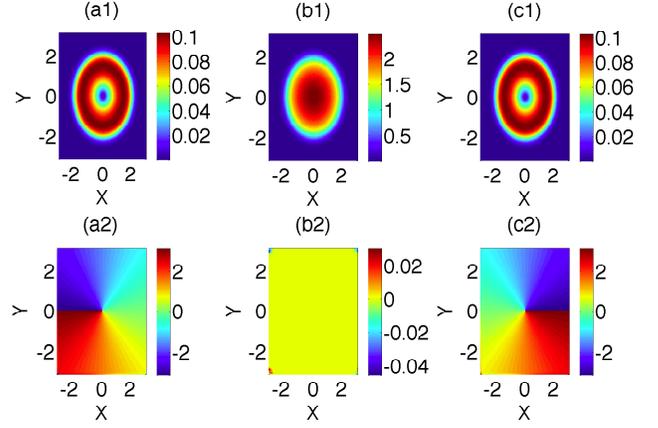}
\caption{(color online) Amplitudes (a1,b1,c1) and phase angles (a2,b2,c2) of the three component wave function
$\psi=(\psi_1,\psi_0,\psi_{-1})^T$ at the ground state of Hamiltonian (\ref{energy}) for a BEC of $^{23}$Na confined
in a 2D harmonic trap. The particle number is $10^6$, the frequency of the trap is $2\pi\times42$ Hz, and the dimensionless SOC strength is $\kappa=0.04$. The units of the $X$ and $Y$ axes are $a_{\rm h}$.}
\label{wf}
\end{center}
\end{figure}

We find that when the SOC is weak ($\kappa\ll 1$), the ground state wave function
has the form
\begin{equation}\label{ground}
\psi=\left(\begin{array}{c}
\chi_{1}(r)e^{-i\phi}\\
\chi_{0}(r)\\
\chi_{-1}(r)e^{i\phi}
\end{array}\right),
\end{equation}
with $\chi_{1}(r)=-\chi_{-1}(r)$ and all $\chi_i$ real. The ground state is
shown in Fig. \ref{wf}. Such a ground state consists of an anti-vortex
in the $m=1$ component and a vortex in the $m=-1$ component.
The $m=0$ component does not carry angular momentum. Since $|\psi_1|=|\psi_{-1}|$,
the net mass current vanishes.

The wave function in Eq. (\ref{ground}) can be understood in the single particle level.
In terms of the ladder operators of spin and angular momentum, the SOC term reads
\begin{equation}
\mathcal{H}_{\rm soc}=\frac{\gamma\sqrt{M\hbar\omega}}{2}\left[S_+\left(\hat{a}_R-\hat{a}_L^{\dagger}
\right)+S_-\left(\hat{a}_R^{\dagger}-\hat{a}_L\right)\right],
\end{equation}
where $S_{\pm}$ is the ladder operator of spin, and $\hat{a}_{L(R)}^{\dagger}$ is the creation operator of the
left (right) circular quanta \cite{cohen}. When the SOC is very weak ($\kappa\ll 1$), its effect can be accounted
for in a perturbative way. From the ground state $\Psi^{(0)}=|0,0\rangle$, the first order correction to the
wave function for small $\gamma$ is given by
\begin{align}
\Psi^{(1)}=&\frac{\gamma\sqrt{M\hbar\omega}}{2\hbar\omega}\left(-S_+\hat{a}_L^{\dagger}
+S_-\hat{a}_R^{\dagger}\right)|0,0\rangle \nonumber \\
=&\frac{\kappa}{2}\left(-|1,-1\rangle+|-1,1\rangle\right),
\end{align}
where $|m_s,m_o\rangle$ denotes a state with spin quantum number $m_s$ and orbital magnetic quantum number
$m_o$. One immediately sees that $\psi_1$ has angular momentum $-\hbar$ and $\psi_{-1}$ has angular momentum $\hbar$.
Besides, the amplitudes of both $\psi_1$ and $\psi_{-1}$ are proportional to $\kappa$.

There exits a continuity equation for spin density and spin current, which  is
\begin{equation}
\frac{d}{d t}\left(\psi^{\dagger}{\bf S}_\mu\psi\right)+\nabla\cdot\mathbf{J}_{\mu}^{s}=0.
\end{equation}
The spin current density tensor $\mathbf{J}_{\mu}^{s}$ ($\mu=x, y, z$ denotes the spin component) is defined as \cite{aref1,aref2}
\begin{align}
\label{def}
\mathbf{J}_{\mu}^{s}=&\frac{1}{2}\left\{\psi^{\dagger}S_\mu{\bf v}\psi+{\rm c.c.}\right\} \nonumber \\
 =&\frac{1}{2}\left\{\sum_{m,n,l}\psi_m^*\left(S_\mu\right)_{mn}{\bf v}_{nl}\psi_l+{\rm c.c.}\right\},
\end{align}
where
\begin{equation}
{\bf v}_{nl}=\frac{{\bf p}}{M}+\gamma\left(\hat{z}\times{\bf S}_{nl}\right),
\end{equation}
and c.c. means the complex conjugate. The second part in ${\bf v}_{nl}$  is
induced by the SOC.

\begin{figure}
\begin{center}
\includegraphics[width=8.5cm]{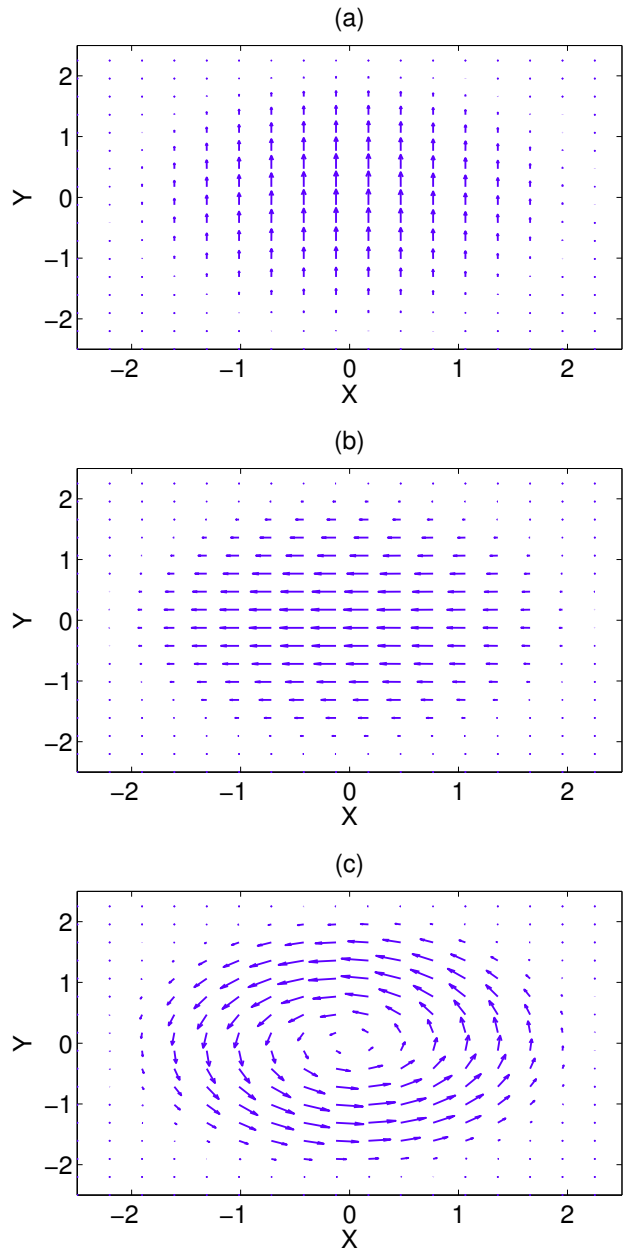}
\caption{(color online) Distribution of the spin current densities $\mathbf{J}_{x}^{s}$ (a), $\mathbf{J}_{y}^{s}$ (b) and
$\mathbf{J}_z^s$ (c)
of the ground state shown in Fig. \ref{wf}. The length of the arrows represents the
strength of the spin current. The arrow length of different subfigures is not to scale. $\kappa=0.04$.
The units of the $X$ and $Y$ axes are $a_{\rm h}$.}
\label{current}
\end{center}
\end{figure}

By the definition in Eq. (\ref{def}), the spin current density carried by
the ground state (\ref{ground}) is
\begin{eqnarray}
\label{currentdensity}
\mathbf{J}_{x}^{s}=&\gamma\sin2\phi|\psi_1|^2\hat{x}+\gamma\left(|\psi_0|^2+2|\psi_1|^2\sin^2\phi\right)\hat{y}, \nonumber\\
\mathbf{J}_{y}^{s}=&-\gamma\left(|\psi_0|^2+2|\psi_1|^2\cos^2\phi\right)\hat{x}-\gamma\sin2\phi|\psi_1|^2\hat{y}, \nonumber\\
\mathbf{J}_{z}^{s}
=&\left(-\frac{2\hbar|\psi_{1}|^{2}}{Mr}+\sqrt{2}\gamma|\psi_1\psi_0|\right)\hat{\phi}.
\end{eqnarray}
From both analytical and numerical results of the wave function, $|\psi_1|\ll|\psi_0|$, so $\mathbf{J}_{x}^{s}$
roughly points in the $y$ direction, while $\mathbf{J}_{y}^{s}$ almost points in the $-x$ direction.
$\mathbf{J}_{z}^{s}$
represents a flow whose amplitude has rotational symmetry.
 From the numerical results shown in
Fig. \ref{current}, we see that $\mathbf{J}_{z}^{s}$ is a counter-clockwise flow.
The amplitudes of $\mathbf{J}_{x}^{s}$ and $\mathbf{J}_{y}^{s}$ are of the same order, both proportional to $\kappa$,
while that of $\mathbf{J}_{z}^{s}$, proportional to $\kappa^2$, is much smaller.
It is evident that the state in Eq. (\ref{ground}) carries no mass current and only pure spin current. Since the spin current is in the ground state, it
must be stable.  In this way,  we have realized a superfluid of pure spin current,
or a pure spin super-current.

\subsection{Experimental schemes}\label{expe}
In this subsection, we propose the experimental schemes to generate and detect
the pure spin currents discussed in subsections \ref{planar} and \ref{circular}.

The planar pure spin current can be easily generated. By  applying a magnetic field gradient,
the two components $m=1,-1$  will be accelerated in opposite directions and a pure
spin current is generated as done in Refs. \cite{engels,hamner}.  To stabilize this
spin current, one needs to generate the quadratic  Zeeman effect. We apply an  oscillating magnetic field $B\sin\omega t$ with the frequency $\omega$ being much larger than
the characteristic frequency of the condensate, e.g., the chemical potential $\mu$.
The time averaging removes the linear Zeeman effect; only the quadratic Zeeman effect remains.  The coefficient of the quadratic  Zeeman effect from the second-order
perturbation theory is given by $\lambda=\left(g\mu_{\rm B} B\right)^2/\Delta E_{\rm hf}$, where $g$ is the Land\'e $g$-factor of the atom, $\mu_{\rm B}$ is the Bohr magneton, and $\Delta E_{\rm hf}$ is the hyperfine energy splitting \cite{ueda}. For the $F=2$ manifold of $^{87}$Rb, $\Delta E_{\rm hf}<0$, so the coefficient of the quadratic Zeeman effect is negative.

The circular flow  in subsection \ref{circular} may find prospective realizations in
two different systems: cold atoms and exciton BEC. In cold atoms, we consider a
system consisting of a BEC of $^{23}$Na confined in a pancake trap, where the
confinement in the $z$ direction is so tight  that one can treat the system effectively
as two dimensional. The SOC can be induced by two different methods.
One is by the exertion of a strong external electric field ${\bf E}$ in the $z$ direction. Due to the relativistic effect,
the magnetic moment of the atom will experience a weak SOC,
where the strength $\gamma=g\mu_B|{\bf E}|/Mc^2$. Here $M$ is the atomic mass
and $c$ is the speed of light.
For weak SOC (small $\gamma$), the fraction of atoms in the $m=1,-1$ components
is proportional to $\gamma^2$.
For an experimentally observable fraction of atoms, e.g., $0.1\%$ of $10^6$ atoms,
using the typical parameters of $^{23}$Na BEC,
the estimated electric field is of the same order of magnitude as the vacuum breakdown field. For atoms with smaller mass or larger magnetic moment, the required electric field can be lowered. Another method of realizing SOC
is to exploit the atom laser interaction, where strong SOC can be
created in principle \cite{socreview}.
In exciton BEC systems, as the effective mass of exciton is much smaller than that of atom, the required
electric field is four to five orders of magnitude smaller, which is quite feasible in experiments \cite{exciton1,exciton2,exciton3,exciton4}.

The vortex and anti-vortex in the $m=1,-1$ components can be detected by the method
of time of flight. First one can split the three spin components with the 
Stern-Gerlach effect. The appearance of vortex or anti-vortex in the $m=1,-1$ components is signaled by a ring structure in the time of flight image. After a sufficiently long time of expansion, the ring structure should be clearly visible.

\section{summary}\label{summary}
In summary, we have studied the superfluidity of three kinds of unconventional superfluids,
which show distinct features from a uniform spinless superfluid. The periodic superfluid
may suffer a new type of instability, the dynamical instability, absent in homogeneous case;
the spin-orbit coupled superfluid has a critical velocity dependent on the reference frame,
a new phenomenon compared with all previous Galilean invariant superfluids;
the superfluid of a pure spin current, though scarcely stable in previous studies,
can be stabilized by the quadratic Zeeman effect and SOC in planar and circular geometry,
respectively. These new superfluids significantly enrich the physics of bosonic superfluids.

With the rapid advances in cold atom physics and other fields, the family 
of superfluids is expanding with the addition of more and more novel superfluids.
Previous study has greatly deepened and enriched our understanding of 
superfluidity, but we believe physics more exciting and beyond 
the scope of our current understanding remains to be discovered in the future. 

\acknowledgments
This work is supported by the NBRP of China (2013CB921903,2012CB921300) and the NSF of China (11274024,11334001,11429402).

\end{CJK*}  
\end{document}